\documentclass{llncs}

\usepackage{times}
\usepackage{epsfig}
\usepackage{graphicx}
\usepackage{amsmath}
\usepackage{amssymb}
\usepackage{caption}
\usepackage{color}
\usepackage{subfigure}
\usepackage{epstopdf}
\usepackage{setspace}
\usepackage{url}
\usepackage{cite}

\providecommand{\keywords}[1]
{
  \small    
  \textbf{\textit{Keywords---}} #1
}
\begin{document}

\title{No-reference Image Denoising Quality Assessment}

\author{Si Lu}

\institute{Portland State University}
\maketitle

\begin{abstract}

A wide variety of image denoising methods are available now. However, the performance of a denoising algorithm often depends on individual input noisy images as well as its parameter setting. In this paper, we present a no-reference image denoising quality assessment method that can be used to select for an input noisy image the right denoising algorithm with the optimal parameter setting. This is a challenging task as no ground truth is available. This paper presents a data-driven approach to learn to predict image denoising quality. Our method is based on the observation that while individual existing quality metrics and denoising models alone cannot robustly rank denoising results, they often complement each other. We accordingly design denoising quality features based on these existing metrics and models and then use Random Forests Regression to aggregate them into a more powerful unified metric. Our experiments on images with various types and levels of noise show that our no-reference denoising quality assessment method significantly outperforms the state-of-the-art quality metrics. This paper also provides a method that leverages our quality assessment method to automatically tune the parameter settings of a denoising algorithm for an input noisy image to produce an optimal denoising result.

\end{abstract}

\keywords{image denoising, quality assessment, random forests regression}

\section{Introduction}
\label{sec:intro}

\begin{figure*}
\begin{footnotesize}
\begin{tabular}{ccccc}

    \hspace{-0.in}\includegraphics[width=.190\textwidth]{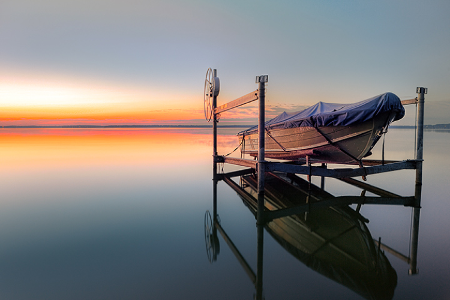} & \hspace{-0.in}\includegraphics[width=.190\textwidth]{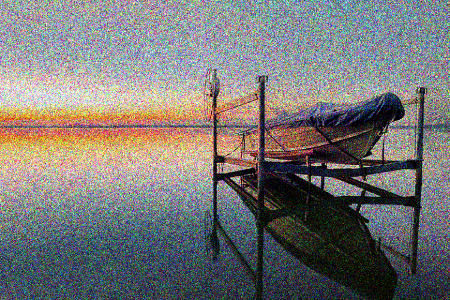} &\hspace{-0.in}\includegraphics[width=.190\textwidth]{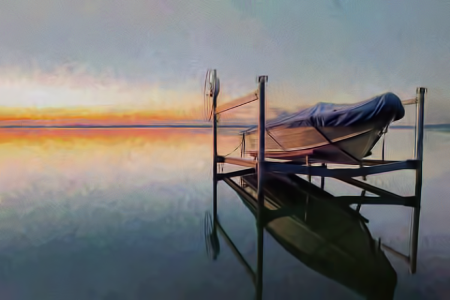} & \hspace{-0.in}\includegraphics[width=.190\textwidth]{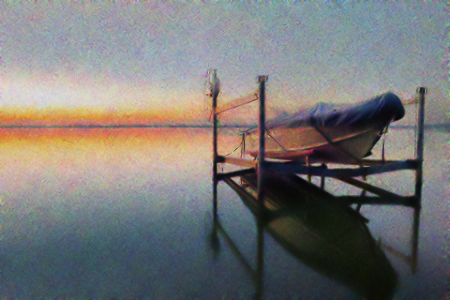}&\hspace{-0.in}\includegraphics[width=.190
    \textwidth]{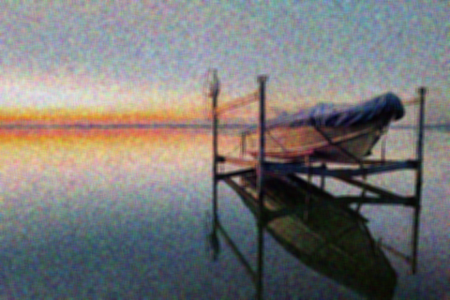}\vspace{0.05in}\\
	
    \hspace{0.in}Ground truth & \hspace{-0.in}Noisy image&\hspace{-0.in} BM3D & \hspace{-0.in}GE & \hspace{-0.in}NLM\\
    
    \hspace{0.in} & \hspace{-0.in} & \hspace{-0.in} PSNR=26.50&\hspace{-0.in}PSNR=23.33&\hspace{-0.in} PSNR=23.17\\
	
	\hspace{-0.in}\includegraphics[width=.190\textwidth]{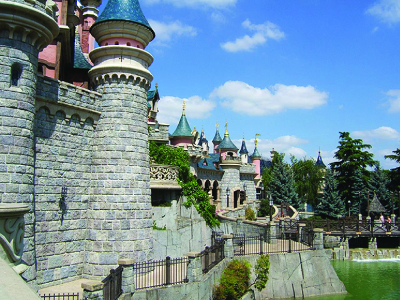} & \hspace{-0.in}\includegraphics[width=.190\textwidth]{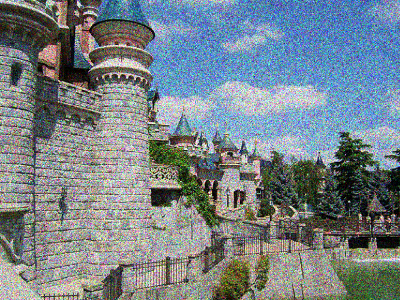} &\hspace{-0.in}\includegraphics[width=.190\textwidth]{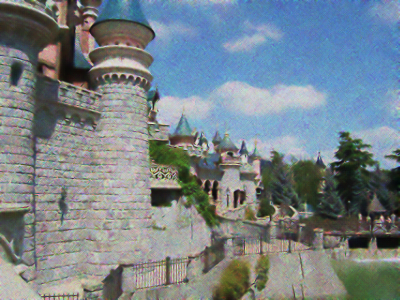} & \hspace{-0.in}\includegraphics[width=.190\textwidth]{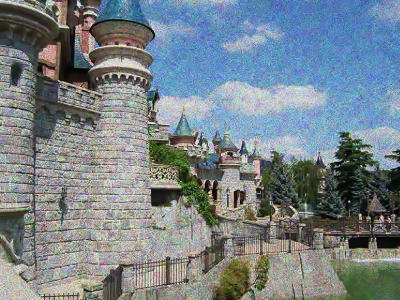}&\hspace{-0.in}\includegraphics[width=.190\textwidth]{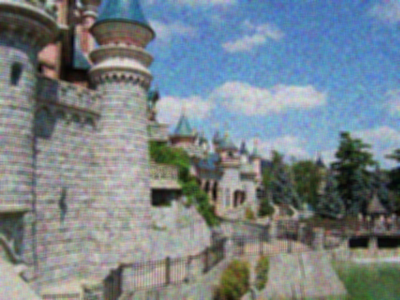}\vspace{0.03in}\\
	
    \hspace{0.in}Ground truth& \hspace{-0.in}Noisy image&\hspace{-0.in} GE&\hspace{-0.in}BM3D&\hspace{-0.in}NLM\\
    
    \hspace{0.in}& \hspace{-0.in} & \hspace{-0.in} SSIM=0.4901&\hspace{-0.in} SSIM=0.4767&\hspace{-0.in} SSIM=0.3959
\end{tabular}
\end{footnotesize}

\caption{Denoising quality ranking. The performance of image denoising algorithms varies over different input images. Our method reliably ranks the denoising results without ground truth. In each row, the denoising results from three methods BM3D~\cite{bm3d}, NLM~\cite{buades:nlm}, and GE~\cite{chen:pg} are listed from left to right according to the rank predicted by our method (from high to low). Our ranking is consistent with the ground truth ranking according to the PSNR values (top row) or SSIM values (bottom row) of the denoising results. Please zoom in to appreciate the quality difference.\label{fig:motiv}}

\end{figure*}

Denoising is one of the basic image processing techniques. A wide range of image denoising methods are now available~\cite{bm3d,buades:nlm,chen:external,elad:sparse,wnnm,ddid,patch-comp-denoising,inter-extern-comb,Roth:foe,bilateral,xu:pgpd,cidweb,internal-stat,across-scale-denoise,epll}. While research on image denoising has been progressing quickly these years, each individual denoising method has its own advantages and disadvantages. No single method works consistently best on different input images, as shown in Figure~\ref{fig:motiv}. Moreover, a denoising method often needs to have its parameter(s) tuned to achieve an optimal denoising result for an input image.



The goal of this paper is to develop an denoising quality assessment method that can be used to select for an input noisy image the right denoising algorithm with the optimal parameter setting. This is a challenging task as no ground-truth image is available for the popular quality metrics like PSNR and SSIM~\cite{wang:ssim} to predict denoising quality. No-reference image quality metrics can be applied to image denoising quality assessment~\cite{Chen_2014_CVPR,mittal:brisque,bovik:biqi,bovik:blinds,tang:lniqa,ye:fln,gu2015using,li2011image}; however, they are designed to measure the overall image quality and do not focus on image denoising quality. Aesthetic quality assessment metrics~\cite{rapid,tang2013content,tian2015query,luo2008photo,Xiaoou_03} have also been proposed. Some other methods aim to do visual quality assessment for screen content images~\cite{SCI_VQA} or tone-mapped images~\cite{tone_map_vqa}. These methods, however, mainly focus on overall image quality/aesthetics assessment. They are not designed to evaluate the quality of (multiple) denoising results. Our work is most relevant to the recent methods for no-reference image denoising quality assessment~\cite{kong:ni,zhu:qmetric}. These two methods measure denoising quality according to some rules or models and have shown promising results. However, when these underlying individual rules or models are not valid for some images, these methods sometimes cannot reliably predict denoising quality.

In this paper, we present a data-driven approach to predict image denoising quality in terms of PSNR and SSIM with no ground truth. The key challenge is to design and extract features from the input noisy image and the denoising result that can be used to learn a denoising quality prediction model. We address this challenge based on two observations. First, although many image denoising methods do not directly target at developing denoising quality metrics, they are developed based on some understanding or model of denoising quality and aim to produce the denoising results that maximize the denoising quality. Accordingly, our method borrows the underlying rules or quality measurement in these denoising methods as well as existing quality metrics to design denoising quality features. Second, while each of existing denoising quality metrics and the implicit quality measurements in various denoising methods alone cannot work robustly, they often complement each other. Therefore, our method aggregates all these features using Random Forests Regression and trains a robust model for image denoising quality prediction that significantly outperforms individual metrics. We further develop a method that uses our model to automatically and efficiently tune the parameter settings for a denoising method to obtain the optimal denoising result. 

The main contribution of this paper is a robust no-reference denoising quality assessment method. This paper shows how individual denoising quality assessment metrics and individual denoising methods can be leveraged to robustly predict image denoising quality without ground-truth. While this paper does not provide a new image denoising algorithm, we provide a way for users to make good use of the large number of denoising algorithms developed by the community over decades. Our experiments show that our method significantly outperforms the recent state-of-the-art no-reference metrics Q~\cite{zhu:qmetric} and SC~\cite{kong:ni} by 0.37 and 0.61 in terms of Kendall's $\tau$ coefficient ($\tau\in$[-1,1]) for image denoising quality ranking. This paper also provides an automatic method to efficiently tune the parameter setting for an image denoising method using our metric. Finally, this paper builds a large image denoising benchmark, which will be shared with the community.

\section{Image Denoising Quality Metric}
\label{sec:method}

In this section, we first discuss what makes a good image denoising result and describe features to measure image denoising quality in Section~\ref{sec:feature}. We then introduce our denoising quality ranking method in Section~\ref{sec:ranking}. We finally describe a method that can automatically tune the parameter setting of a method to optimize the denoising result for an input noisy image in Section~\ref{sec:para}.

\subsection{Denoising Quality Feature Design}
\label{sec:feature}

There has been a literature on image denoising. We observe that many of these denoising methods aim to optimize certain denoising quality metrics defined according to various observations and rules. We hypothesize that although the denoising quality metrics in existing denoising methods~\cite{ksvd,zuo:ghp} and some dedicated denoising quality assessments~\cite{kong:ni,zhu:qmetric} alone are still inadequate for denoising quality prediction, they complement each other. We accordingly design features to encode denoising quality based on these existing quality metrics. In this section, we use $I$ and $\hat{I}$ to denote an input noisy image and its denoised image, respectively. We define the extracted noise map (image) as the difference between $\hat{I}$ and $I$ and denote it as $I_n=\hat{I}-I$.

\begin{figure} [ht]
\begin{small}
\begin{tabular}{ccc}

\hspace{-0.00in}\includegraphics[width=.32\textwidth]{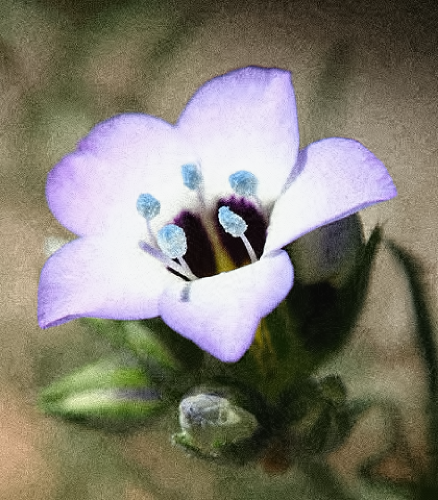} & \hspace{-0.00in}\includegraphics[width=.32\textwidth]{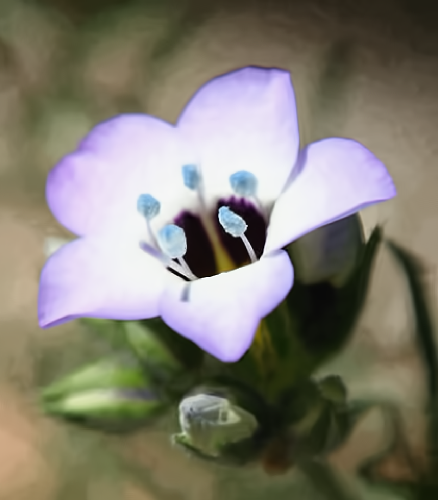} & \hspace{-0.00in}\includegraphics[width=.32\textwidth]{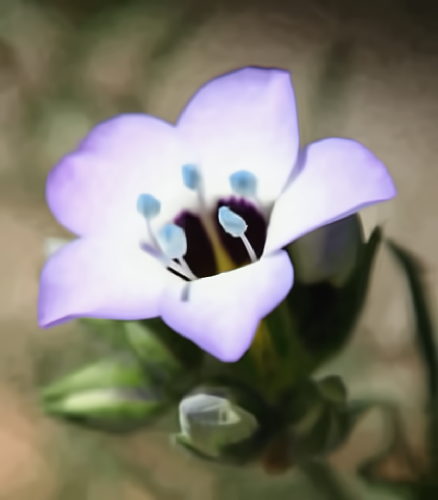}\\
 \hspace{-0.00in}(a) Under-smoothed result& \hspace{-0.00in}(b) Optimal denoising result& \hspace{-0.00in}(c) Over-smoothed result\\
 \hspace{-0.00in}SS=0.88& \hspace{-0.00in}SS=0.28& \hspace{-0.00in}SS=0.20
    \end{tabular}
    \end{small} 
    \caption{$SS$ values for different denoising results.
    \label{fig:SS_Example}}
\end{figure}

\begin{figure}[htb]
\begin{center}
\begin{small}
\begin{tabular}{c}
\hspace{-0.075in}\includegraphics[width=1.0\textwidth]{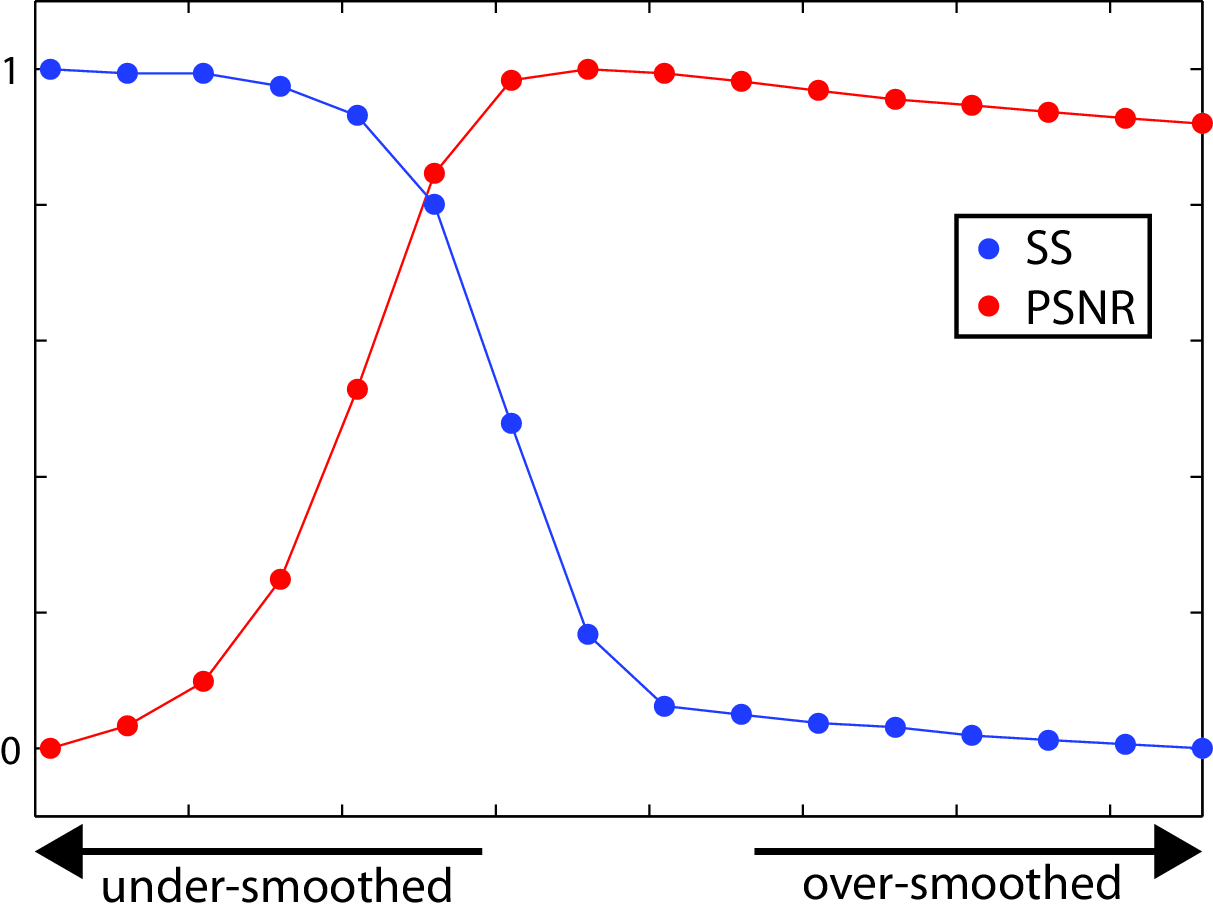} \\
    \end{tabular}
    \end{small} 
    \caption{$SS$ values for denoising results with various smoothness. All denoising results are generated by BM3D with different parameters. Note that the PSNR values in this figure are normalized to (0,1)
    \label{fig:SS_Stat}}
\end{center}
\end{figure}

\noindent\textbf{Local Self-similarity.} Local patches in an image often share similar patterns. Many patch-based image denoising algorithms~\cite{ksvd,bm3d} find similar patches and denoise these patches simultaneously. It has been our observation that the property of self-similarity often manifests more significantly in noise-free images than noisy images. We accordingly compute the self-similarity between patches in a denoised image to measure image denoising quality. Specifically, we first uniformly divide a denoised image into {$k \times k$} image patches with $k=15$. We then vectorize each patch into a column vector and assemble all these columns into a patch matrix. We apply Singular Value Decomposition(SVD) to the patch matrix and get a singular value vector $S$.
We finally measure the self-similarity value $SS$ as the minimal number of the largest singular values such that the sum of these largest singular values is at least $\alpha$ of the sum of all the singular values.
\begin{eqnarray}
SS=\frac{t}{|S|}, \mbox{where }\sum_{i=1}^{t}s_i>=\alpha\sum_{i=1}^{|S|}s_i
\end{eqnarray}
To make our measurement robust against the choice of $\alpha$, we try three different $\alpha$ values, namely 0.97, 0.98, and 0.99, and accordingly compute three local self-similarity values.

Figure~\ref{fig:SS_Example} and figure~\ref{fig:SS_Stat} show the local self-similarity values of different denoising results of the same input image generated by BM3D with different parameter settings. This example shows that the local self-similarity value correlates well with the smoothness of the denoising results instead of the actual denoising quality. We address this problem with two methods. First, we use a data-driven approach to find the right decision boundary values w.r.t. this local self-similarity feature to identify the right feature value range of a good denoising result. Second, as discussed before, we do not expect that a single feature will be sufficient and we jointly use other features to measure image denoising quality. {As shown in our experiment in Section~\ref{sec:exp}, although this feature alone is insufficient, it contributes to the overall performance of image denoising quality prediction.}

\begin{figure} [htb]
\begin{small}
\begin{tabular}{ccc}
  
  \hspace{-0,in}\includegraphics[width=.32\textwidth]{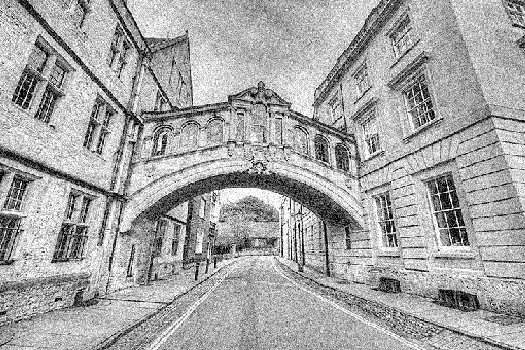} &
  \includegraphics[width=.32\textwidth]{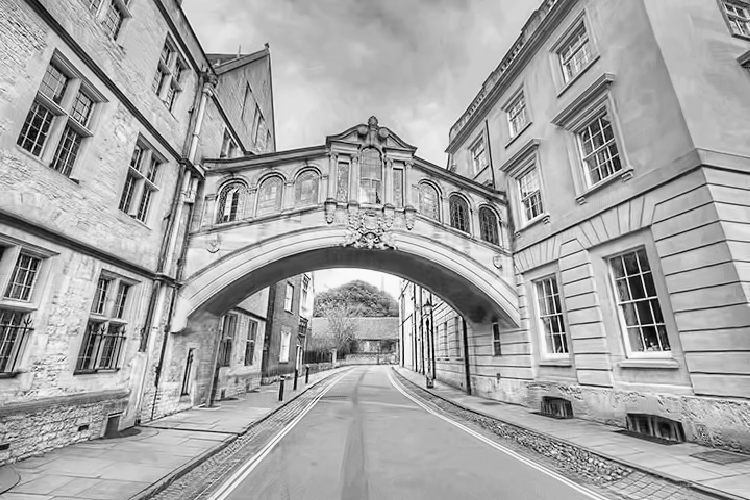}     &
  \includegraphics[width=.32\textwidth]{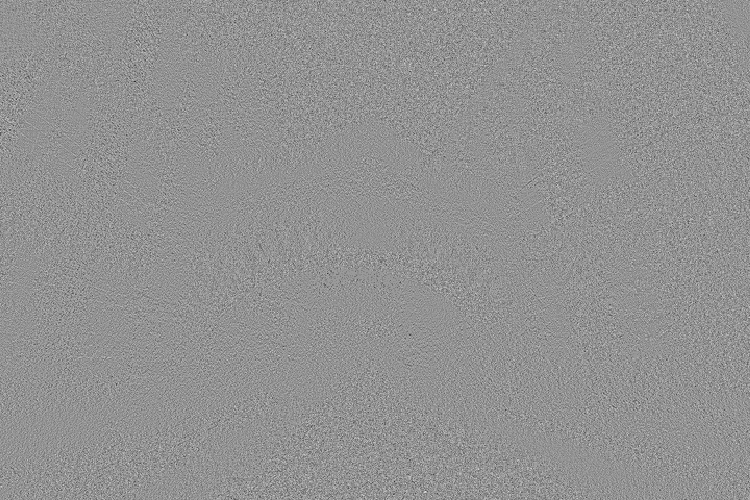}     \\
  (a) Noisy image.  &  (b) Good result, PSNR=27.2   &  (c)  $I_n$ of (b), $SR_1=10.7$\\
  
  \hspace{-0,in}\includegraphics[width=.32\textwidth]{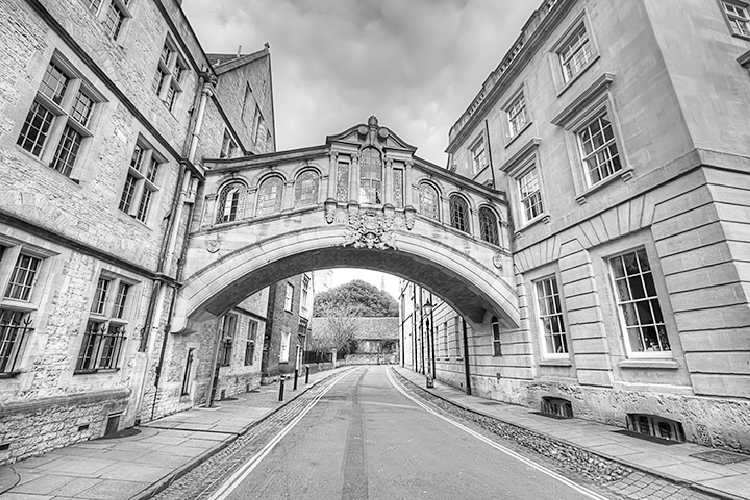} &
  \includegraphics[width=.32\textwidth]{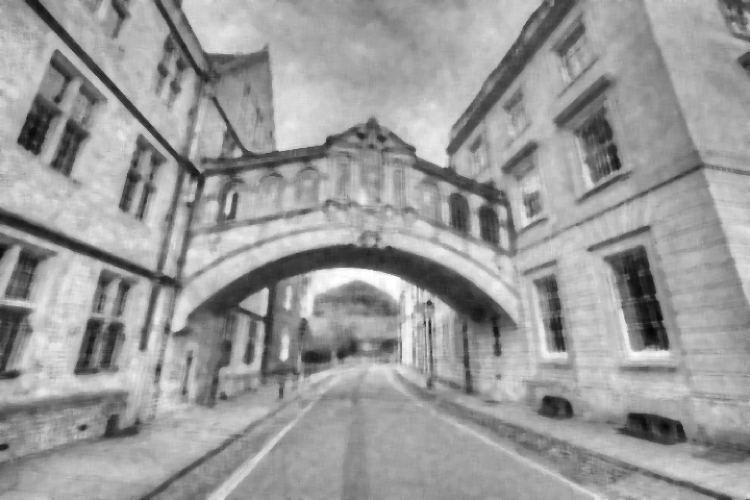}     &
  \includegraphics[width=.32\textwidth]{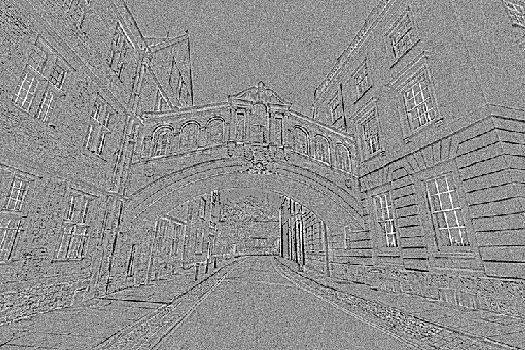}     \\
  
  (d) Clean image.  &  (e) Bad result, PSNR=17.8   &  (f)  $I_n$ of (e), $SR_2=22.0$

\end{tabular}
\end{small} 
\caption{$SR$ for two different denoising results. \label{fig:SR_Example}}
\end{figure}

\vspace{0.1in}
\noindent\textbf{Structural Residual.} The original image structure in an input noisy image should be preserved as much as possible during denoising~\cite{buades:nlm}. Thus, the noise layer $I_n$, which is the difference between the input noisy image and the denoised result, should contain as little structure residual as possible. Our method estimates the residual structure as a feature to measure image denoising quality. Our method first uses the method from Chen et.al.~\cite{chen:sr} to compute a structure residual map to capture the amount of structures in the noisy layer $I_n$. Specifically, the structural residual at pixel $p$ is computed as a weighted average of a number of nearby and ``similar'' noise samples in $I_n$.

\begin{equation}
I^r_n(p)=\sum_{q\in\Omega}w(p,q,\sigma_{d},\sigma_{c},\sigma_{s})I_n(q)
\end{equation}
\noindent where $\Omega$ is the neighborhood of pixel $p$. $w(p,q,\sigma_{d},\sigma_{c},\sigma_{s})$ is a weight function that computes how the noise value at pixel $q$ contributes to the residual structure computed at pixel $p$. $\sigma_{d}$, $\sigma_{c}$ and $\sigma_{s}$ are the standard deviation parameters that control the relative impacts of the spatial distance, color distance, and image structure difference.

After we obtain the structural residual map $I^r_n$, we compute the structure residual feature $SR$ as follows.
\begin {equation}
SR=\sqrt{\frac{\sum_p{I^r_n(p)^2}}{N}}
\end {equation}
\noindent where $N$ is the total number of pixels in the image.

\begin{figure} [htb]
\centering
  \begin{minipage}[t]{0.44\linewidth}
    \centering
    \includegraphics[width=1.0\textwidth]{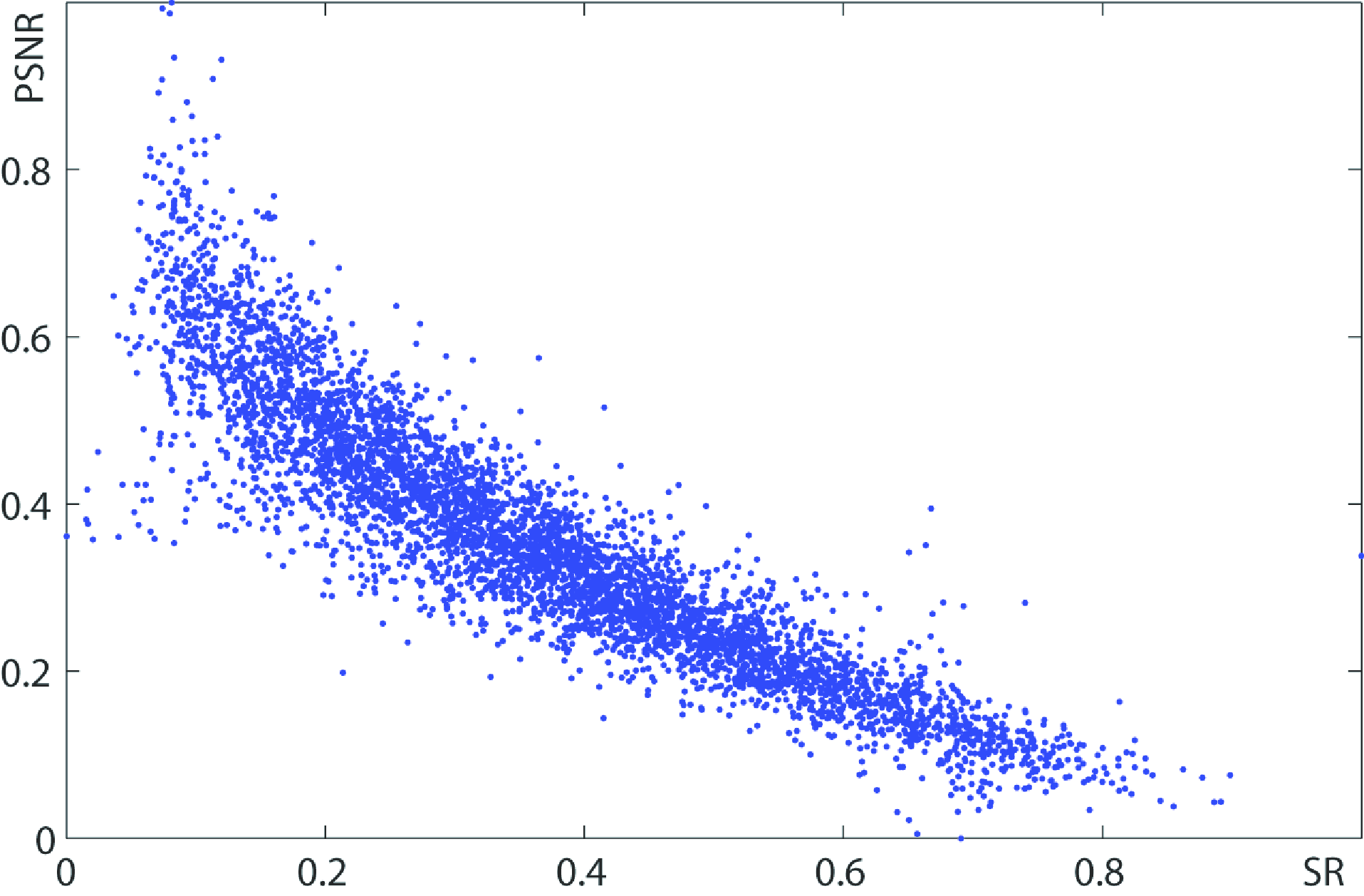}
    \caption{$SR$-PSNR di5stribution.}
    \label{fig:SR_Stat}
  \end{minipage}
  \hspace{8ex}
  \begin{minipage}[t]{0.44\linewidth}
    \includegraphics[width=1.0\textwidth]{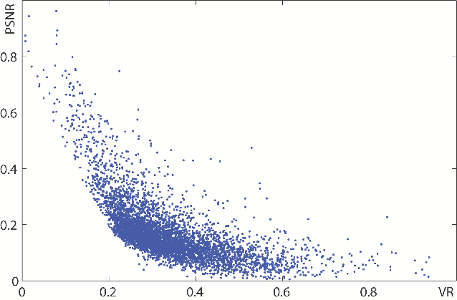}
    \caption{$VR$-PSNR distribution.}
    \label{fig:VR_Stat}
  \end{minipage}
\end{figure}

To capture structural residuals at different image scales, we use three sets of parameters ($SR_1:\sigma_d=\sigma_s=1.0, \sigma_c=4.0, SR_2:\sigma_d=\sigma_s=4.0, \sigma_c=10.0, SR_3:\sigma_d=\sigma_s=10.0, \sigma_c=30.0$), as suggested in~\cite{chen:sr}, and accordingly estimate three structure residual feature values for each denoising result. As shown in Figure~\ref{fig:SR_Stat}, the $SR$ value of a denoising result is statistically negative correlated with the image denoising quality. Figure~\ref{fig:SR_Example} shows the $SR$ values of both a good denoising result and a bad denoising result. There is nearly no structural residual in the good denoising result.

\begin{figure} [hb]
\centering
\begin{small}
  \begin{tabular}{cccc}
    \hspace{-0.02in}\includegraphics[width=.245\textwidth]{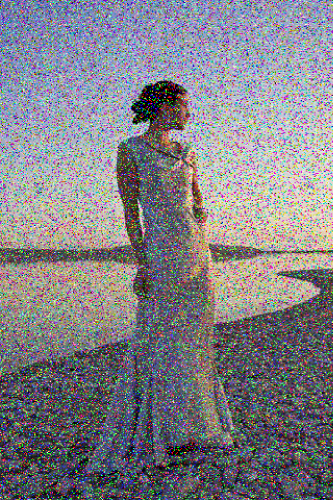}&
	\hspace{-0.02in}\includegraphics[width=.245\textwidth]{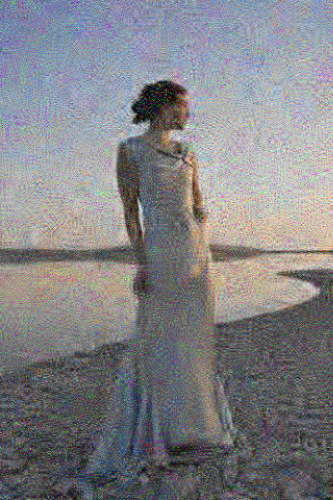}&
	\hspace{-0.02in}\includegraphics[width=.245\textwidth]{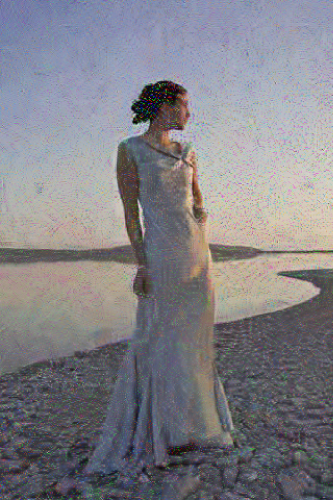}&
	\hspace{-0.02in}\includegraphics[width=.245\textwidth]{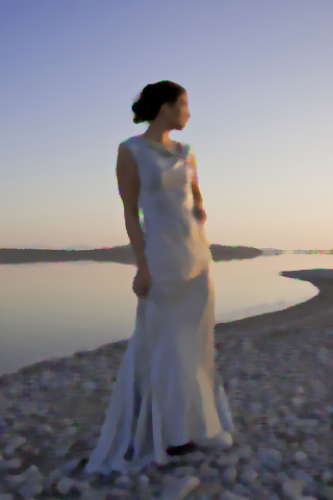}\\

	{Noisy input}  &  SGM=4.59    &  SGM=2.22    &  SGM=0.18   \\
                   &  PSNR=19.15  &  PSNR=22.05  &  PSNR=27.01

  \end{tabular}
\end{small}
\caption{$SGM$ values of different denoising results. \label{fig:SGM_Example}}
\end{figure}

\vspace{0.1in}
\noindent\textbf{Small Gradient Magnitude.}  According to Liu et.al.~\cite{liu:debiqa}, gradients of small magnitude often correspond to noise in flat regions. We follow their approach and use the standard deviation of the $m\%$  smallest non-zero gradient magnitudes in the denoising result to measure the denoising quality. To make this small gradient magnitude feature ($SGM$) robust against the parameter $m$, we set $m=40,50,60$ and compute three corresponding feature values. Figure~\ref{fig:SGM_Example} shows an example of this feature.

\begin{figure} [ht]
  \begin{small}
    \begin{tabular}{cc}
      \hspace{-0.in}\includegraphics[width=.495\textwidth]{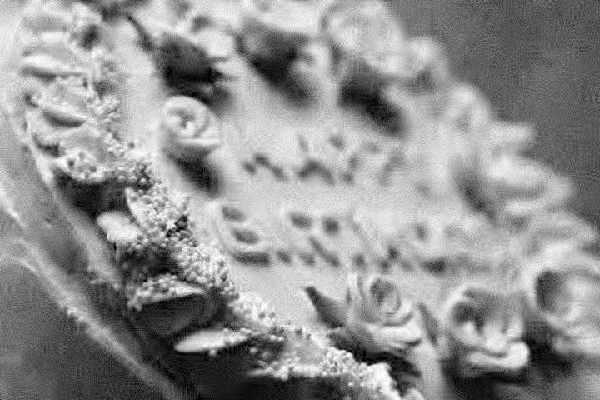}&
      \hspace{-0.in}\includegraphics[width=.495\textwidth]{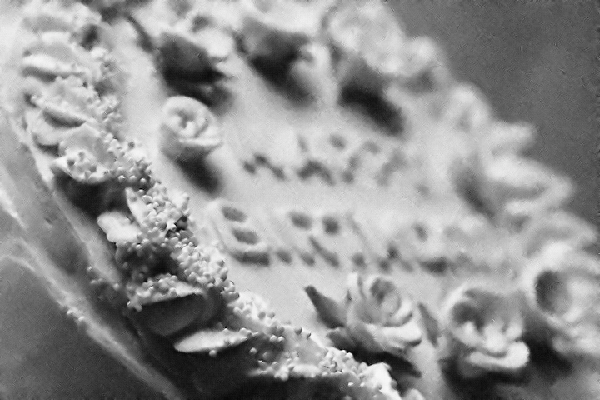}\\
      {SC=0.607, PSNR=26.30}  &  SC=0.692, PSNR=30.03 
    \end{tabular}
  \end{small}
  \caption{$SC$ values for different denoising results. \label{fig:SC_Example}}
\end{figure}

\noindent\textbf{Adaptive Structure Correlation.} The structure similarity between a well denoised image $\hat{I}$ and the input noisy image $I$ should be low in the flat region and high in the highly textured region, while the structure similarity between the resulting noise map $I_n$ and the input noisy image $I$ should be high in the flat region and low in the textured region~\cite{kong:ni}. Then for a good denoising result, the structure similarity map between the denoised image and the input noisy image and the structure similarity map between the resulting noisy map and the input noisy image should be negatively correlated. Our method follows the denoising quality metric in~\cite{kong:ni} and uses the negative correlation measurement between the two structure similarity maps to encode denoising quality as follows.
\begin{equation}
SC=-corr(I_{ss}(\hat{I},I),I_{ss}(I_n,I))
\end{equation}
where $I_{ss}(\cdot)$ computes the structure similarity map between two input images according to the structure similarity metric \emph{SSIM}~\cite{wang:ssim}. To better capture structure correlation among different image scales, we choose three different patch sizes, namely $6\times6$, $8\times8$ and $10\times10$, to compute three feature values. Figure~\ref{fig:SC_Example} shows $SC$ values of denoised images with different qualities.

The performance of the above variation denoising formulation depends on both the parameter $\lambda$ and the selection of the norm operator. We therefore use various combination of values for $\lambda$ and the norm operations, as reported in Table~\ref{tab:DO_para}, and compute multiple feature values accordingly.

\begin{figure} [htb]
\begin{small}
\begin{tabular}{ccc}
  \hspace{-0.0in}\includegraphics[width=.32\textwidth]{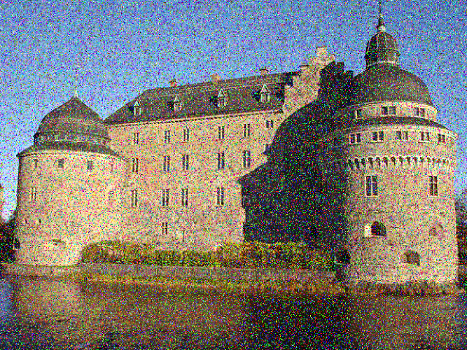} & \hspace{-0.0in}\includegraphics[width=.32\textwidth]{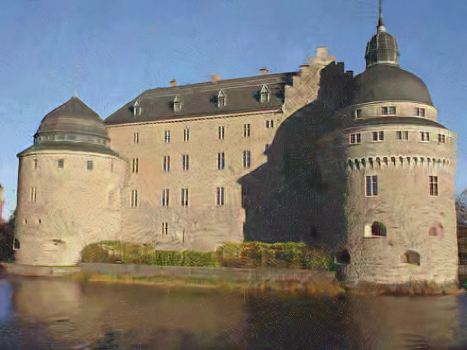} & \hspace{-0.0in}\includegraphics[width=.32\textwidth]{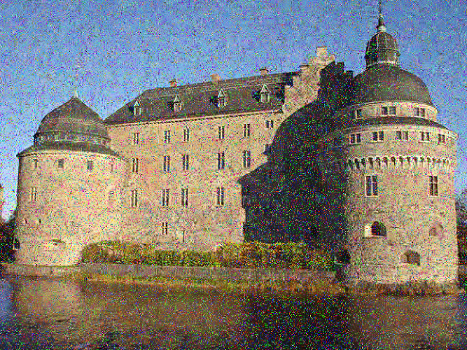} \\
  \hspace{-0.0in}{Noisy input}&\hspace{-0.0in}{$VR$=2.28*e4, PSNR=21.58}&\hspace{-0.0in}{$VR$=2.45*e4, PSNR=15.49}\\
  &\hspace{-0.0in}{Good denoising result}&\hspace{-0.0in}{Bad denoising result}
\end{tabular}
\end{small}
\caption{$VR$ values for different denoising results. \label{fig:VR_Example}}
\end{figure}


\begin{table}[htb]
\centering
\caption {Parameter selection for $VR$. \label{tab:DO_para}}
\begin{tabular}{c|c|c|c|c|c|c}
  \hline
   &$VR_1$& $VR_2$& $VR_3$ & $VR_4$& $VR_5$&$VR_6$\\
  \hline
  $l_d$ & $\ell1$  & $\ell1$ &$\ell2$& $\ell2$ & $\ell2$ & $\ell2$  \\
  $l_s$ & $\ell1$  & $\ell1$ &$\ell1$& $\ell1$ & $\ell2$ & $\ell2$  \\
  $\lambda$ &0.5 & 1 &0.5& 1 & 0.5 & 1  \\
  \hline
\end{tabular}
\end{table}	   

\noindent\textbf{Variational Denoising Residual.} A range of variational methods have been developed for image denoising~\cite{tv:denoise,zhang:mlt}. These methods formulate image denoising as an optimization problem with the pixel values in the denoising result as variables. Many of these methods formulate denoising as the following optimization problem.
\begin{equation}
\min_{\hat{I}}\frac{\|I-\hat{I}\|_{l_d}+\lambda \|\nabla\hat{I}\|_{l_s}}{N}
\end{equation}
\noindent where the first term is the data term that encourages the denoising result $\hat{I}$ should be as close to the input image $I$ as possible. The second term is a regularization term that aims to minimize the gradient of the denoising result $\nabla\hat{I}$. $\lambda$ is a parameter. $l_d$ and $l_s$ indicate the norm operator used in each term. There are three popular norm operator combinations used in existing denoising methods, as reported in Table~\ref{tab:DO_para}. $N$ is the number of pixels in the image. Variational methods minimize the above energy function to obtain the denoising result. Therefore, we compute the above energy function value denoted as $VR$ by plugging in the de-noising result to measure the denoising quality. Figure~\ref{fig:VR_Example} shows that $VR$ values are negative correlated with the denoising qualities. Note that when $\ell2$ norm is used, the corresponding term is computed as the square of the norm.

\noindent\textbf{Gradient Histogram Preservation.} A good denoising result should preserve as many image structures as possible. A recent image denoising method from Zuo et.al. first estimates a target image gradient histogram $H_g$ and then aims to find such a denoising result that has a gradient histogram as similar to the target gradient histogram as possible~\cite{zuo:ghp}. Accordingly, our method uses the difference between the gradient histogram of an denoising result $\hat{I}$ and the target gradient histogram to indicate the denoising quality.
\begin{equation}
GH=\|H_g-\hat{H}\|_{\ell1}
\end{equation}
\noindent where $H_g$ is the normalized target gradient histogram estimated using a method from~\cite{zuo:ghp} and $\hat{H}$ is the normalized gradient histogram of the denoising result $\hat{I}$.

\subsection{Denoising Quality Prediction}
\label{sec:ranking}

Given an input noisy image $I$ and its $k$ denoising results $\{\hat{I}_i\}_{i=1,\cdots,k}$, we aim to predict the denoising quality without ground truth. Specifically, given an input noisy image $I$ and its denoising result $\hat{I}$, we model the image denoising quality as a Random Forests Regression ~\cite{rfReg} of the quality feature vector $f_{I, \hat{I}_i}$ extracted from $I$ and $\hat{I}$ using the methods described in Section~\ref{sec:feature}.

\begin{equation}
\label{eq:pairwise}
q_{I,\hat{I}_i} = RFR( f_{I, \hat{I}_i})
\end{equation}

\noindent where $q_{I,\hat{I}_i}$ indicates the denoising quality and $RFR$ is the Random Forests Regression model that is trained using our training dataset. We directly use the output value of this Random Forest Regression model to indicate denoising quality and obtain the final quality ranking according this regression value. For color images, we independently assess the denoising quality of each channel and then averaging them to obtain the final quality assessment results.


\subsection{Automatic Parameter Tuning}
\label{sec:para}

Our denoising quality metric can be used to select a good denoising result that a denoising method can produce. Like previous work~\cite{kong:ni}, we can try all the reasonable algorithm parameters to denoise an input noisy image and use the prediction model to pick the best one. This brute-force approach, however, is time-consuming. Instead, we formulate parameter tuning as the following optimization problem.
\begin{eqnarray}
\theta^*=\arg\max_{\theta}q(\theta)
\end{eqnarray}
\noindent where $q(\cdot)$ is the denoising quality computed according to Equation~\ref{eq:pairwise} with a denoising parameter setting $\theta$. We then use a Gradient Ascent method to find the optimal parameter settings as follows:

\begin{equation}
\label{eq:ParaTuning01}
\theta_{k+1}=\theta_{k}+{\lambda}\nabla q(\theta_k)
\end{equation}

\noindent where $\lambda$ is the step size. $\nabla q(\theta_k)$ is the denoising quality gradient computed using finite difference:

\begin{equation}
    \nabla q(\theta_k)=\frac{q(\theta_k+d\theta)-q(\theta_k-d\theta)}{2d\theta}
\end{equation}

\section{Experiments}
\label{sec:exp}

We developed an image denoising quality assessment benchmark and experimented with our denoising quality ranking method on it. Our experiments compare our method with the state-of-the-art methods $Q$~\cite{zhu:qmetric} and $SC$~\cite{kong:ni}. Below we first describe our image denoising benchmark. We then validate our regression model and evaluate the performance of our method w.r.t. overall quality ranking. We finally evaluate how our method can be used to tune the parameter setting of a denoising algorithm. 

As PSNR and SSIM are two popular metrics that reflect different aspects of image quality assessment, we trained two different models to estimate the denoising quality according to PSN and SSIM, respectively. We then test our method on denoising quality assessment according to both PSNR and SSIM in our experiments.

\subsection{Image Denoising Quality Benchmark}
\label{sec:dataset}

We download 5,000 high-quality images from \emph{Flickr} and use them as the noise-free ground-truth images. These images cover a wide range of scenes and objects, such as landscape, buildings, people, flowers, animals, food, vehicles, etc. For convenience, we downsample these images so that their maximum height is 480 pixels. We then add synthetic noise to each of the original images. We use three types of noise, namely Additive White Gaussian noise, Poisson noise, and Salt \& Pepper noise. For the Gaussian noise, we use three different standard deviation values, namely $\sigma=10$, $20$, and $30$. For the Poisson noise, we use the Poisson random number generation function $poissrnd$ in Matlab to generate noisy images with Poisson noise: $I_n=k\cdot poissrnd(I/k)$ with $k=0.05$, $0.10$ and $0.15$. For the Salt \& Pepper, we add noise with three density levels, namely $d=0.1$, 0.2, and 0.3. Thus, we create 9 noisy images for each input image and obtain $5000\times3\times3=45000$ noisy images in total.

We then denoise each noisy image using seven (7) representative denoising algorithms, namely Gaussian Filter, Bilateral Filter~\cite{bilateral}, Median Filter~\cite{Weiss:fmb}, Non-Local Means~\cite{buades:nlm}, Geodesic denoising~\cite{chen:pg}, DCT denoising~\cite{guo:dct}, and BM3D~\cite{bm3d}. For each algorithm, we choose several different parameter settings to generate denoising results. The numbers of parameter settings for these algorithms are reported in Table~\ref{tab:paranum}. For each noisy image, we have 23 denoising results. For each denoising result, we compute the PSNR and SSIM value and use them as the ground truth labels to train different models for denoising quality assessments.

\begin{table}[htb]
  \begin{center}
  \caption{Numbers of parameter sets of denoising methods.} \label{tab:paranum}
    \begin{tabular}{cccccccc}
      \hline
      Algorithm          & GF & BF & MED & NLM & GE & DCT & BM3D\\
      \# of settings        & 3   & 4   & 3      & 4     & 3   &  3    & 3  \\
      \hline 
  \end{tabular} 
  \end{center}
\end{table}

We partition our benchmark into a training set and a testing set. We randomly select 250 (5\%) out of the 5,000 clean images to generate the training set and the rest to generate the testing set. For each noisy image, we have 23 denoising results. In total, we obtain 250$\times$9$\times$23=51750 denoising results and use them to train the denoising quality predication model. We repeated this random partition 10 times and take the average as the final result.

\subsection{Regression Validation}
\label{sec:regvalid}
\begin{figure} [htb]
  \begin{small}
    \begin{tabular}{cc}
      \includegraphics[width=.495\textwidth]{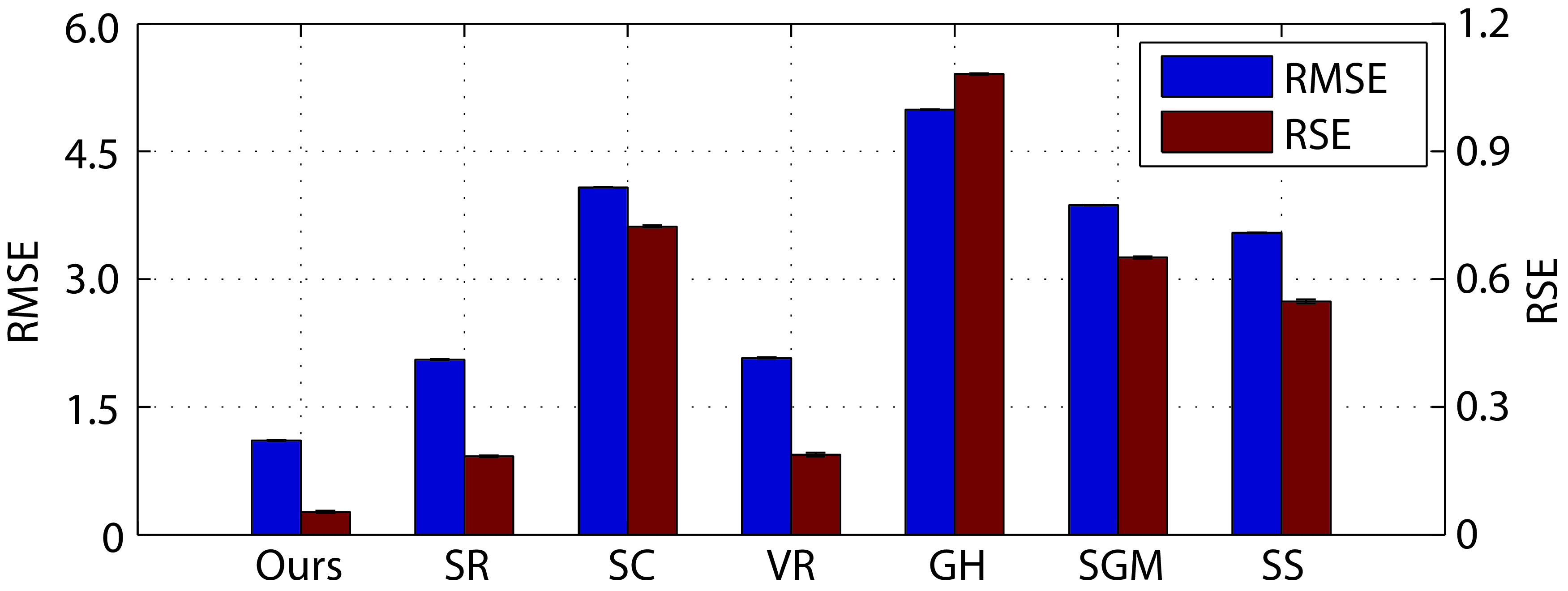}&
      \includegraphics[width=.495\textwidth]{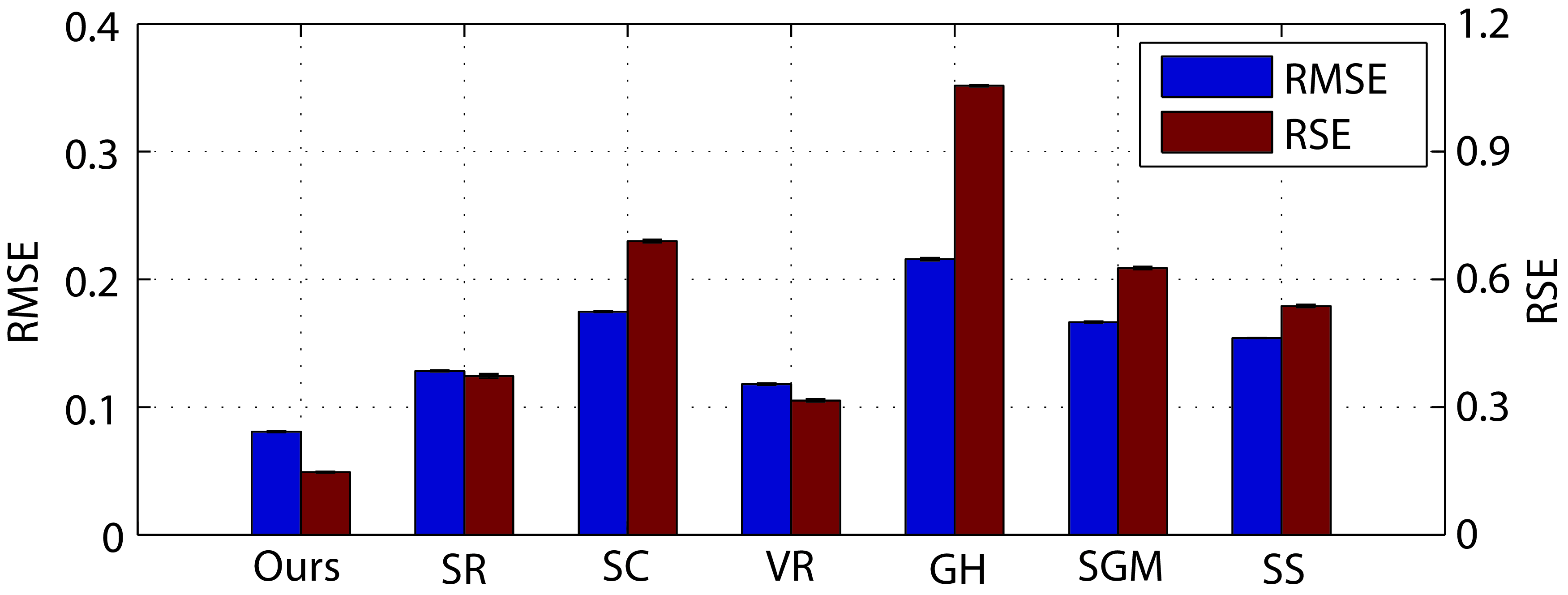}\\
  
      (a) PSNR prediction  &  (b) SSIM prediction 

    \end{tabular}
  \end{small}
  \caption{Regression model evaluation.\label{fig:regValid:rmse}}
\end{figure}

We evaluate our regression model for PSNR/SSIM prediction w.r.t the Root Mean Sqaured Error (RMSE) between our predicted values and the ground truth and the Relative Squared Error (RSE) in Figure~\ref{fig:regValid:rmse}. We also compare our quality prediction using all the features to our method using individual features. This result shows that by aggregating all features together, our method performs significantly better than any individual features.

\begin{figure} [htb]
  \begin{small}
    \begin{tabular}{cc}
      \includegraphics[width=.495\textwidth]{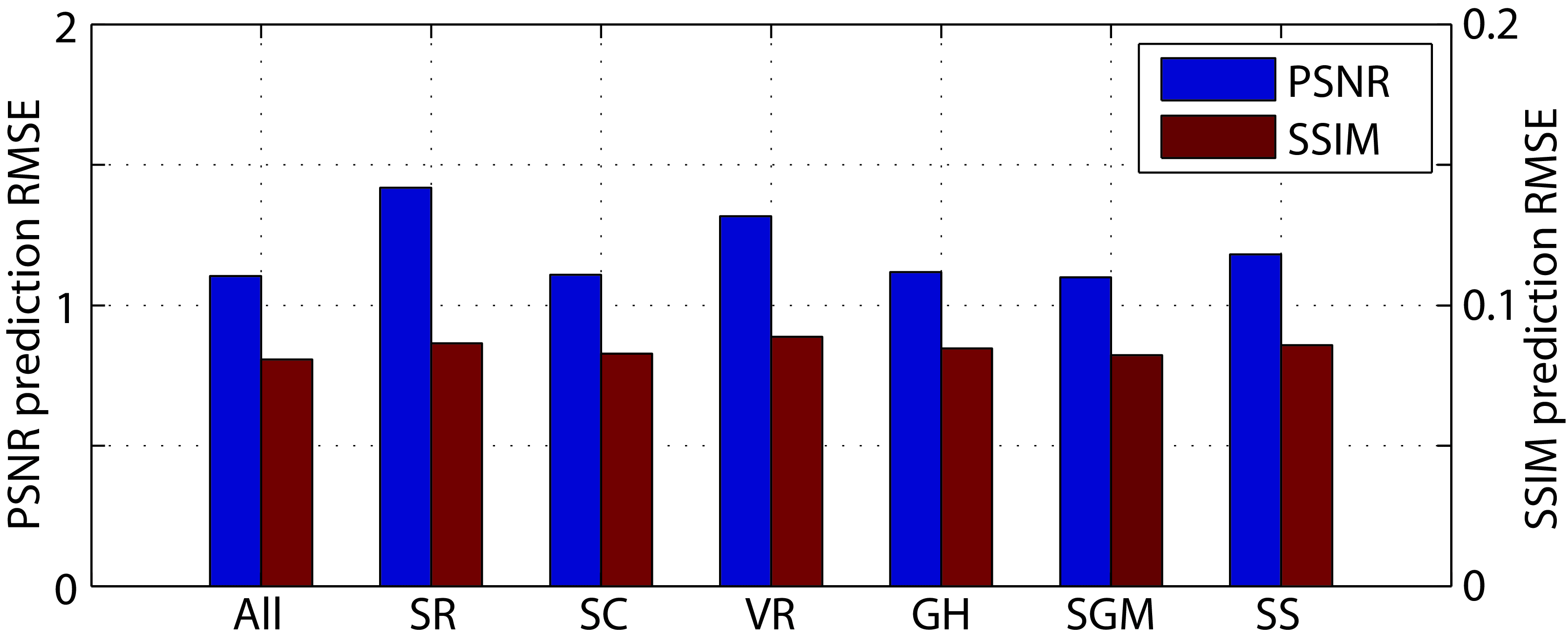}&
      \includegraphics[width=.475\textwidth]{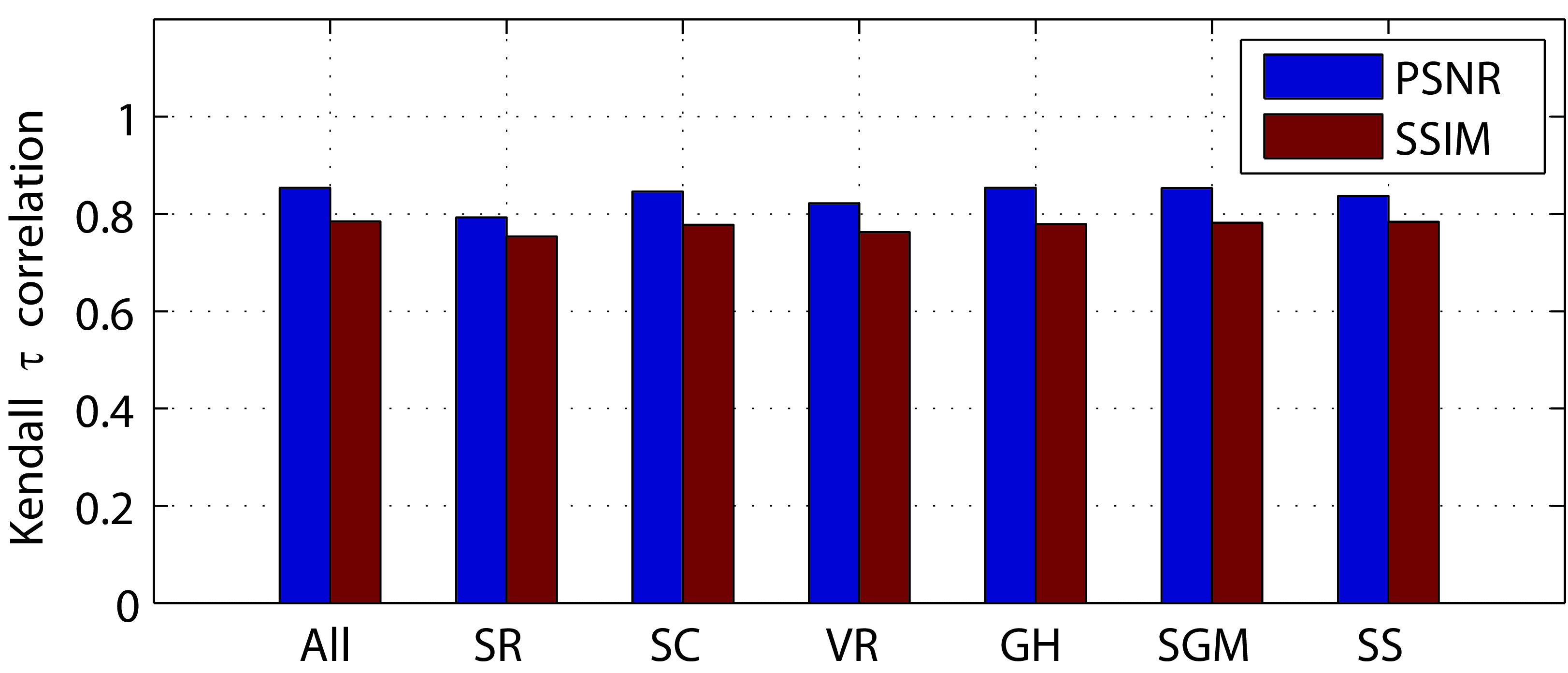}\\
  
      (a) Prediction error  &  (b) Ranking performance

    \end{tabular}
  \end{small}
  \caption{Leaving-one-out experiment results.\label{fig:regValid:leaveOneOut}}
\end{figure}

To further study how each individual feature contributes to final PSNR/SSIM prediction, we leave out that feature and use the rest features train our model and report the results in Figure~\ref{fig:regValid:leaveOneOut} (a). This result shows that removing any single features only slightly downgrades the performance.


\begin{figure} [htb]
\centering
  \begin{small}
    \begin{tabular}{cc}
      \includegraphics[width=.495\textwidth]{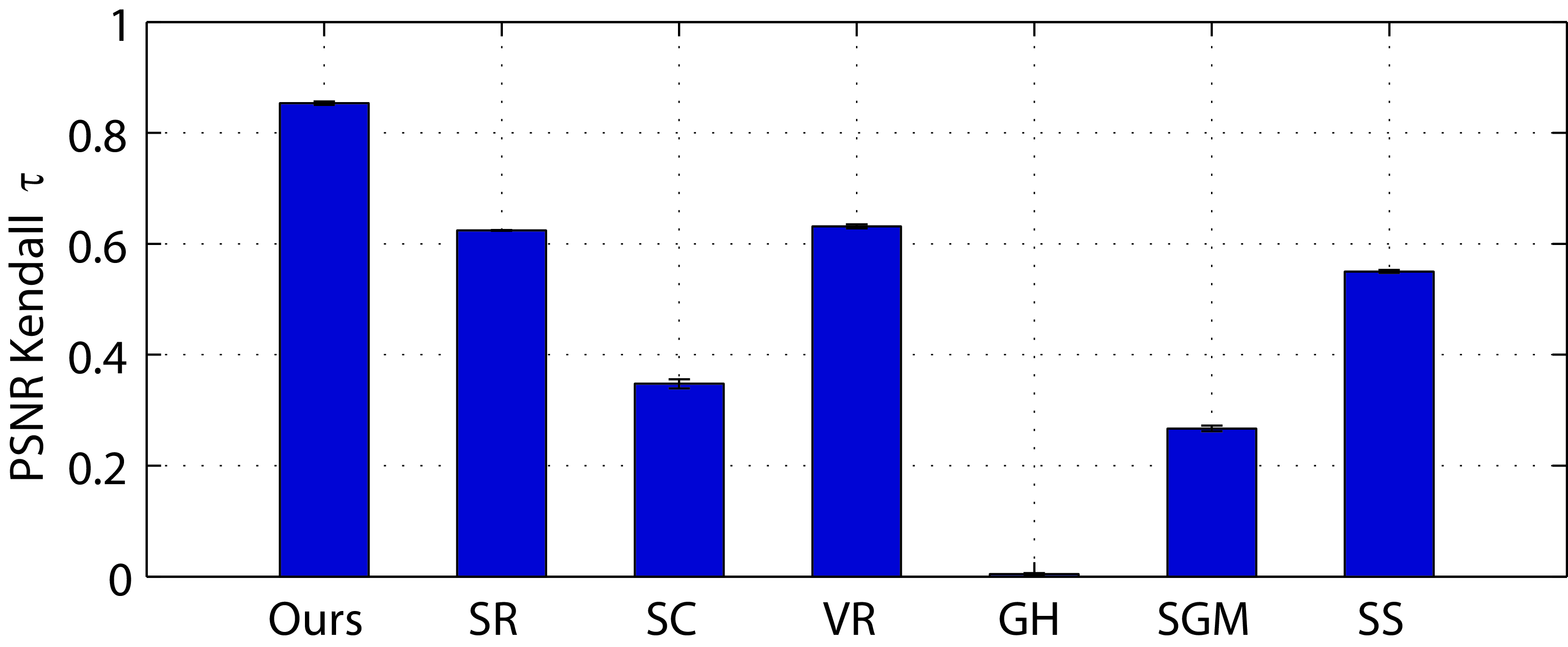}&
      \includegraphics[width=.495\textwidth]{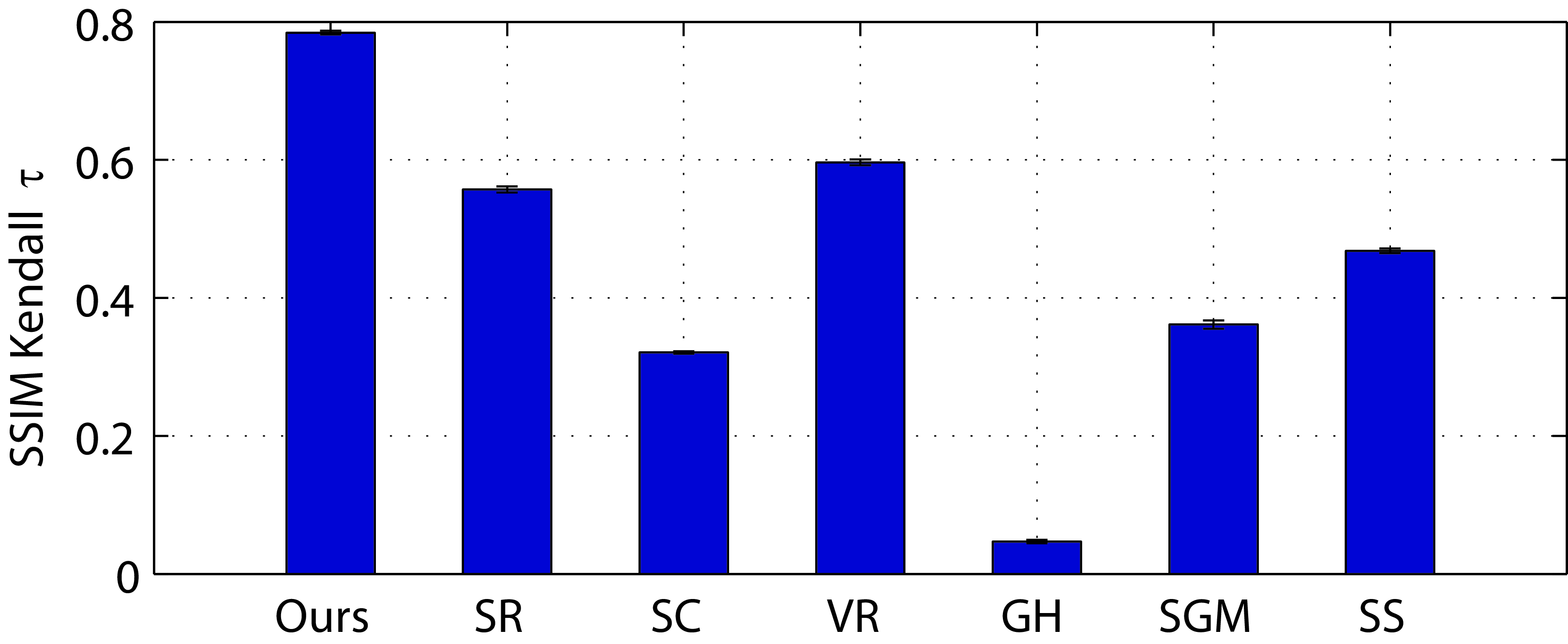}\\
      (a) PSNR prediction  &  (b) SSIM prediction 
    \end{tabular}
  \end{small}
  \caption{All features vs. individual features on denoising quality ranking.\label{fig:pw_dis}}
\end{figure}

\begin{table} [tb]

\centering
\caption{Comparison of our method and existing denoising metrics $Q$, $SC$, $BRISQUE$ and $NIQE$. $\tau_{PSNR}$ and $\tau_{SSIM}$ indicate that the ground truth labels used for training are PSNR and SSIM values, respectively.} \label{tab:dircomp}
\begin{tabular}{cccccc}
  \hline
  Metric               &  Ours & $Q$ & $SC$ &$BRISQUE$ &$NIQE$\\
  $\tau_{PSNR}$        & 0.854                & 0.483              & 0.241               &0.326                    &0.506                \\
  $\tau_{SSIM}$        & 0.784                & 0.453              & 0.191               &0.250                    &0.399                \\
  \hline
\end{tabular}
\hspace{1ex}
\centering
\caption {Performance our method on images on noise levels different from the training set. $\tau_{PSNR}$ and $\tau_{SSIM}$ indicate that the ground truth labels used for training are PSNR and SSIM values, respectively.\label{tab:dircomp_dif_levels}}
\begin{tabular}{cccccc}
  \hline
  Metric       & Ours &$Q$ & $SC$ &$BRISQUE$ &$NIQE$\\
  $\tau_{PSNR}$& 0.681               & 0.436              & 0.239               &0.363                    &0.482                \\
  $\tau_{SSIM}$& 0.507               & 0.488              & 0.200               &0.199                    &0.356                \\
  \hline
\end{tabular}
\end{table}	

\subsection{Evaluation on Denoising Quality Ranking}
\label{sec:exp:pair}

\begin{figure*}[ht]
    \includegraphics[width=0.99\textwidth]{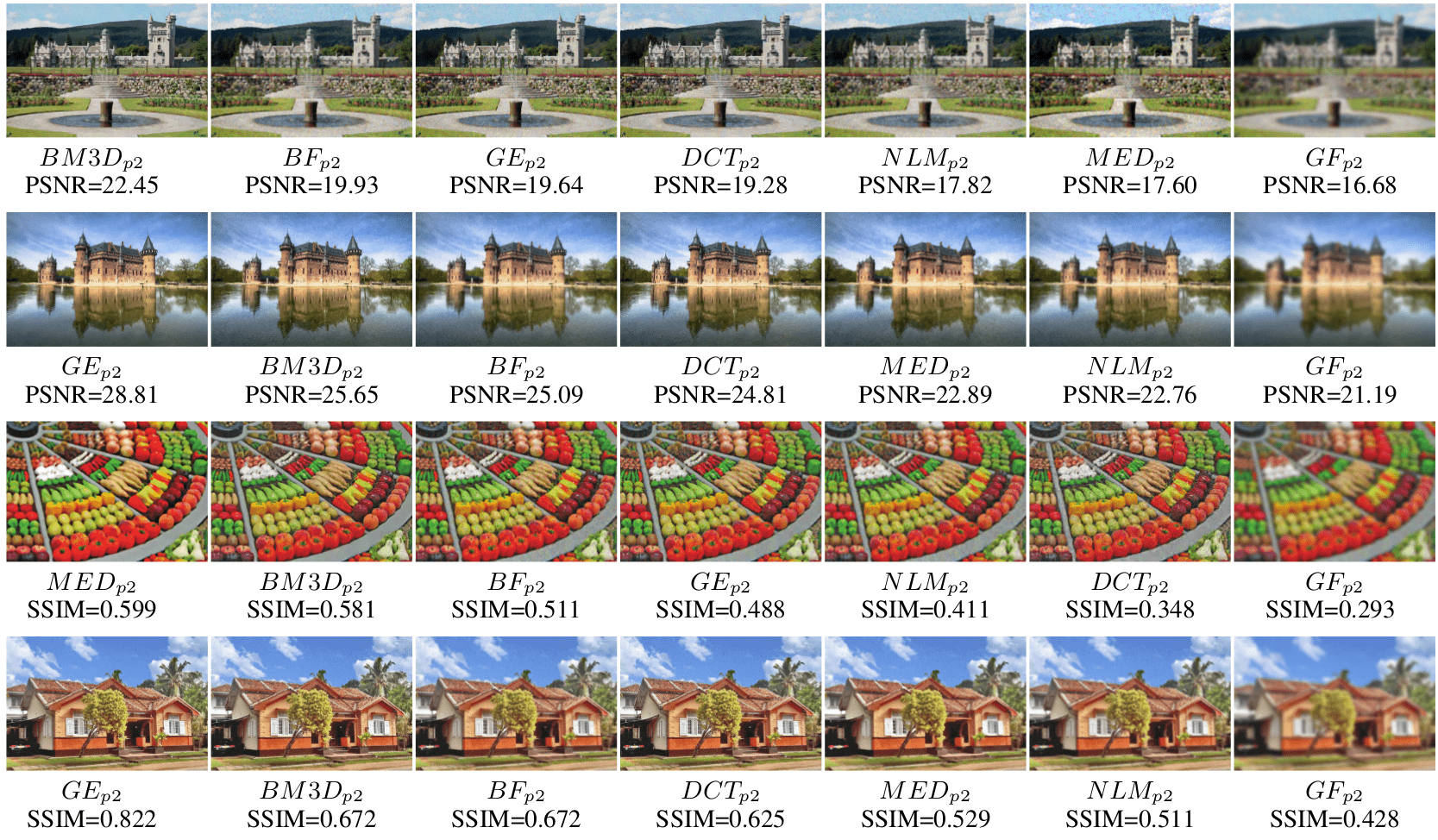}
    \caption{Our method reliably ranks the denoising results in terms of PSNR and SSIM. The denoising results are listed from left to right according to our predicted quality(from high to low). Due to the space limit, we only show 7 denoising results for each example. \label{fig:res}}
\end{figure*}

We adopt the Kendall $\tau$ correlation~\cite{kendall:corr} between our ranking result and the ground truth to evaluate the performance of our ranking method. We directly compare our method with $Q$~\cite{zhu:qmetric}, $SC$~\cite{kong:ni}, $BRISQUE$~\cite{brisque} and $NIQE$~\cite{niqe}in Table~\ref{tab:dircomp}. As can be seen, our approach can better rank denoising results than these state-of-the-art image quality metrics. Figure~\ref{fig:res} shows some denoising quality ranking results of our method. 

To evaluate how our method performs across different noise levels, we select 100 clean images from the testing set and corrupt them with Gaussian noise of $\sigma=15$, $25$, Poisson noise of $k=0.075$, $0.125$ and Salt \& Pepper noise of $d=0.15$, 0.25. Note that images with these noise levels are not used during the training process. We report the performance of our method on these images in Table~\ref{tab:dircomp_dif_levels}. Our method outperforms existing metrics on these noise levels. This result indicates that our method can effectively and robustly estimate image denoising quality across different noise levels.

We compare our method using all the features to our method using individual features in Figure~\ref{fig:pw_dis}. It can be seen that the model trained using all features significantly outperforms models trained using individual features. This result proves our observation that although individual features alone is not enough to correctly estimate the denoising quality, they complement each other well and can thus produce robust quality assessment.

To further evaluate how each individual feature contributes to denoising quality ranking, we leave out that feature and use the rest features to rank denoising results and report the results in Figure~\ref{fig:regValid:leaveOneOut} (b). This leave-one-out test shows that removing any single feature only downgrades the ranking performance very slightly. This indicates that any single feature can almost be replaced by the combination of the rest of the features. In addition, by exploring multiple $SC$ values for training, our method has improved the Kendall $\tau$ ranking performance of the $SC$ metric from $0.239/0.200$ to $0.348/0.321$ for denoising quality assessment according to PSNR$/$SSIM. These tests show the capability of our method in aggregating multiple weak features into a more powerful quality ranking method.

\subsection{Evaluation on Parameter Tuning}
\label{sec:exp:tune}

\begin{figure*}
    \includegraphics[width=0.99\textwidth]{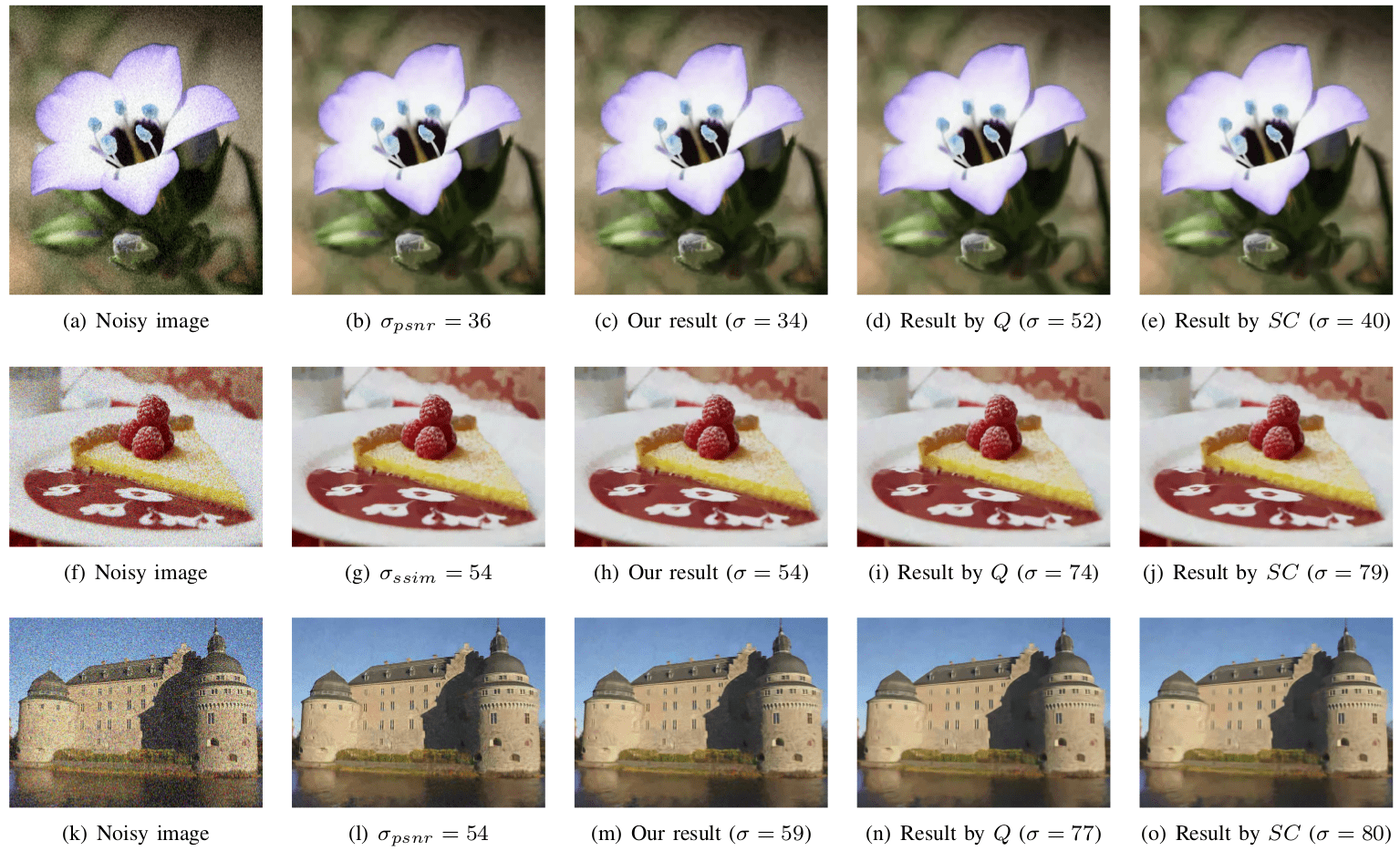}
    \caption{Parameter tuning results for BM3D using our PSNR or SSIM predicting models. Comparing to $Q$ and $SC$, our predicted parameter setting is closer to the  ground truth optimal parameter setting $\sigma_{psnr}$ and $\sigma_{ssim}$.\label{fig:para_predic}}
\end{figure*}

\begin{figure*}
	\subfigure{\includegraphics[width=.49\textwidth]{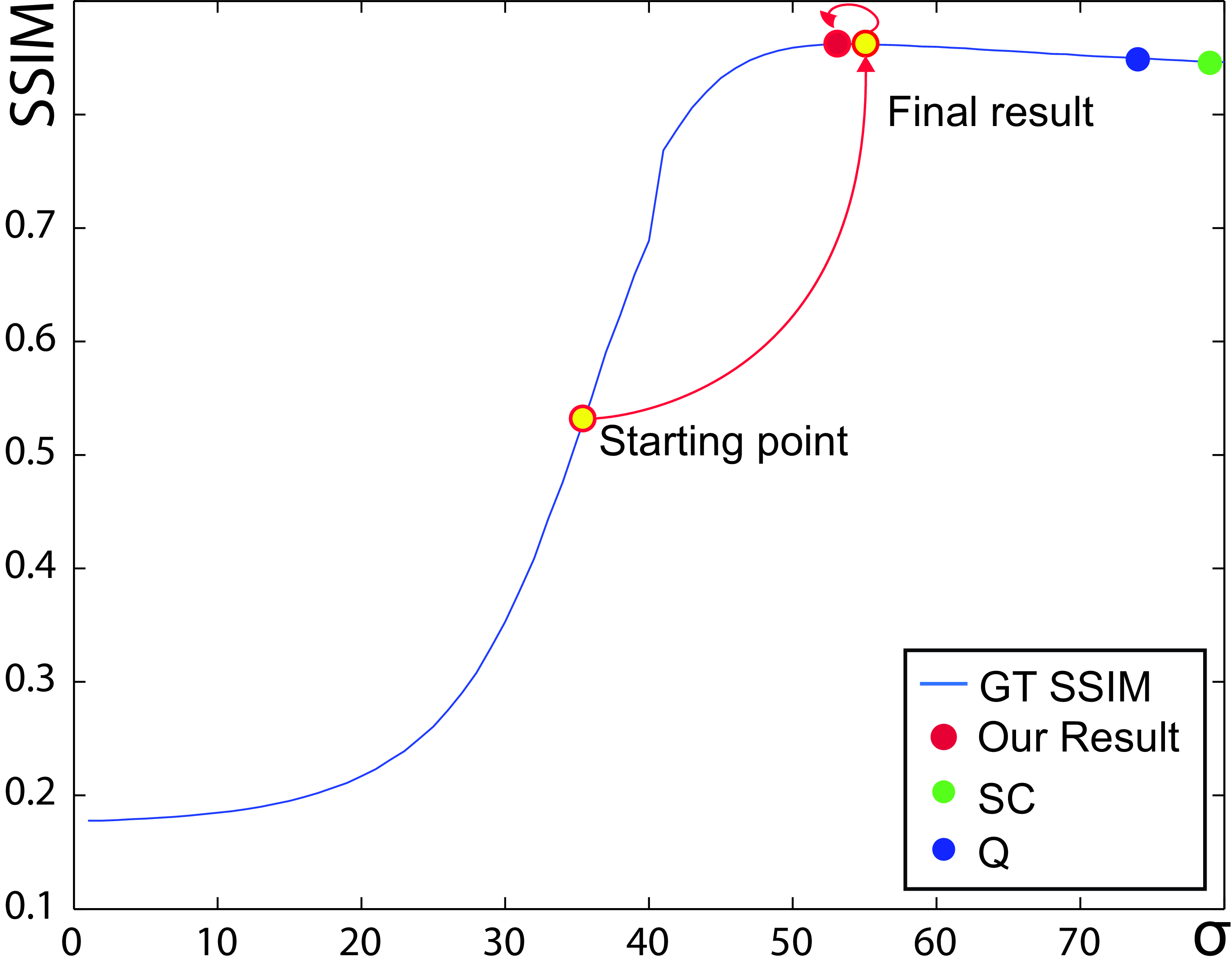}}
    \subfigure{\includegraphics[width=.49\textwidth]{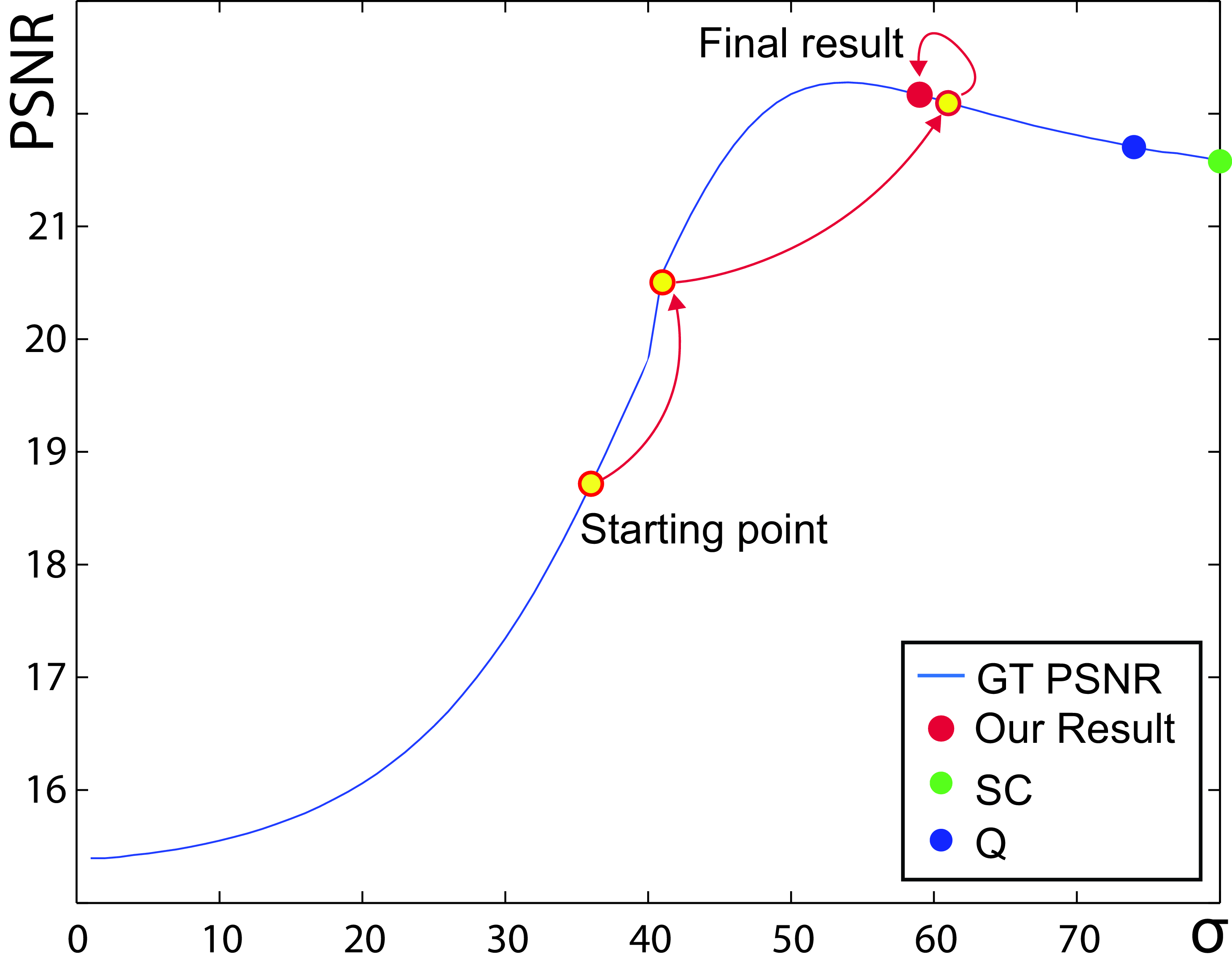}}
    \caption{Parameter tuning process of Figure \ref{fig:para_predic} (f) (k). \label{fig:para_process}}
\end{figure*}

We selected the BM3D~\cite{bm3d} algorithm to evaluate our automatic parameter tuning method. We randomly select 50 noise-free images, add three types of noise to each of them in the same way as described above, and create 150 noisy images. We obtain the ground-truth optimal parameters setting for each noisy image using a brute-force method according to the PSNR and SSIM values. To test the performance of our parameter tuning method, we randomly select half of the noisy images as the training set and the rest half as the testing set. For each noisy image in the training set, we use BM3D to denoise it using 80 different parameter settings, from $\sigma=1$ to $\sigma=80$. We use all images in the training set to train dedicated predicting models for BM3D in purpose of predicting PSNR and SSIM. We then use these models in our automatic parameter tuning algorithm to select an optimal parameter setting for the noisy image in the testing set. The step size $\lambda$ is set to 2 for the PSNR prediction model. Since the SSIM value ranges from 0 to 1, which is smaller than that of PSNR values, we use a larger $\lambda$ value (20) for the SSIM prediction model. For the initial guess $\sigma_{0}$, we find that simply selecting the center of the parameter searching space works well for all our testing images.

A straightforward way to evaluate the performance of our method is to compute the error between our estimated algorithm parameters to the ground-truth optimal parameters. However, this parameter error is difficult to intuitively understand the performance of our parameter tuning algorithm in finding optimal parameters to produce the optimal denoising result. Therefore, we evaluate our method w.r.t. the quality of the final denoising result using the estimated parameter setting. Specifically, we compute the mean PSNR/SSIM difference($Diff_{PSNR}$ and $Diff_{SSIM}$) between the denoising results using the estimated parameter setting and the result using the optimal parameter setting. We compare our parameter tuning method with the state-of-the-art denoising metrics $Q$ and $SC$ in Table~\ref{tab:tune}. For $Q$ and $SC$, we use brute-force search to directly select their estimated parameter settings according to the $Q$ and $SC$ values. It can be seen that our approach outperforms $Q$ and $SC$ on optimal parameter tuning. The denoising results produced with the parameters estimated by our method are closer to the results produced with the ground-truth parameters. On average, the mean $Diff_{PSNR}$ value is 0.426 and the mean $Diff_{SSIM}$ is 0.033. Figure \ref{fig:para_predic} shows three examples of our parameter tuning method.

\begin{table} [tb]
\centering
\centering
\caption{Parameter Tuning Performance.} \label{tab:tune}
\begin{tabular}{p{20mm}p{10mm}p{10mm}p{10mm}p{10mm}p{10mm}p{10mm}}
  \hline
&  \multicolumn{2}{c}{Our method} & \multicolumn{2}{c}{$Q$} & \multicolumn{2}{c}{$SC$} \\
  \hline
                &  mean          & std            & mean   & std   & mean  &  std  \\ 
$Diff_{PSNR}$    & \textbf{0.426} & \textbf{0.673} & 1.165  & 1.817 & 1.650 & 2.659 \\
$Diff_{SSIM}$    & \textbf{0.033} & \textbf{0.047} & 0.066  & 0.140 & 0.097 & 0.183 \\
  \hline
\end{tabular}
\hspace{3ex}
\centering
\caption {Computation cost of our tuning algorithm. \label{tab:speed_para}}
\begin{tabular}{ccc}
  \hline
                    & \multicolumn{2}{c}{Number of iterations} \\
  \hline                 
                    &  mean               & std                         \\ 
  PSNR prediction   & 2.53                & 0.98                        \\
  SSIM prediction   & 2.84                & 1.56                        \\
  \hline
\end{tabular}
\end{table}

To evaluate the computation cost of our parameter tuning, we record the number of iterations during the tuning process and report the results in Table \ref {tab:speed_para}. For each noisy input image, our method converges in about less than 3 iterations. It takes about 5 seconds for our method to predict the quality of one denoised image. Thus, comparing to brute-force searching that requires denoising the image with all possible parameter settings, our method can save a lot of computation cost on denoising process while introducing only a small amount of computation on quality prediction. Figure \ref{fig:para_process} shows our parameter searching process of the noisy image from Figure \ref{fig:para_predic} (f) and (k). This example shows that our method effectively and efficiently reaches a parameter setting that is very close to the real optimal selected according to the Ground Truth (GT) PSNR/SSIM.

\bibliographystyle{spmpsci}
\bibliography{egbib}
\end{document}